\documentclass[twocolumn,times,tighten]{aastex631}

\newcommand*{\toreferee}{\color{black}} 

\usepackage{amsmath}
\usepackage{physics}
\usepackage{bbold}
\usepackage{soul}
\usepackage{booktabs} 
\usepackage{multirow} 
\usepackage{amsmath}  
\usepackage{ulem}

\graphicspath{{./}{figures/}}

\begin{document}



\title{Tracing magnetic field in super-{\toreferee Alfv{\'e}nic} turbulence with Gradient Technique}

\correspondingauthor{Ka Wai Ho}
\email{kho33@wisc.edu / kawaiho@lanl.gov}

\author[0000-0003-3328-6300]{Ka Wai Ho}
\affiliation{Theoretical Division, Los Alamos National Laboratory}
\affiliation{Department of Astronomy, University of Wisconsin-Madison, Madison, WI, 53706, USA}

\author[0000-0002-7336-6674]{A. Lazarian}
\affiliation{Department of Astronomy, University of Wisconsin-Madison, Madison, WI, 53706, USA}
\email{lazarian@astro.wisc.edu}

\begin{abstract}
Super-{\toreferee Alfv{\'e}nic} turbulence is important for many astrophysical objects, particularly galaxy clusters. In this paper, we explore the accuracy of Synchrotron Intensity Gradients (SIGs) and X-ray intensity gradients to map magnetic fields in super-{\toreferee Alfv{\'e}nic} turbulence for a set of astrophysically relevant parameters of turbulent driving. Analyzing our synthetic observations, we report a good accuracy for both techniques. Our results are suggestive that other types of Gradient Technique (GT) can be successfully employed to trace magnetic fields within super-{\toreferee Alfv{\'e}nic sub-sonic} turbulence.
\end{abstract}


\section{Introduction}

Magnetic fields and turbulence are ubiquitous in astrophysical settings. 
MHD turbulence plays an important role on various scales, from millions of parsecs for intracluster medium (ICM) to hundreds of parsecs in the interstellar medium and parsecs, parsecs and astronomical units for star formation, and hundreds of kilometers for stellar winds. The magnetization of the media is extremely important for understanding key astrophysical problems, e.g., the problem of star formation \citep{MO07,2004RvMP...76..125M} and cosmic ray propagation and acceleration.\citep{J66,YL08}.

Studies of astrophysical magnetic fields are challenging. All magnetic measurements rely on the effects of the magnetic field on the media. The Zeeman technique is the most direct way of measuring the strength of the line-of-sight component of the magnetic field that relies on splitting atomic levels (\cite{2010MNRAS.402L..64C} and ref. therein). However, it is a very challenging approach regarding data requirements and observational time. Measuring synchrotron polarization relies on the interaction of relativistic electrons with the magnetic field \cite{Beck1996}. The resulting polarization is perpendicular to the magnetic field. Faraday rotation can affect the synchrotron polarization, which complicates the interpretation of the signal. In addition, Faraday depolarization decreases the polarization signal, especially at low frequencies \citep{LP16}. 
At the same time, the Faraday rotation from point sources, e.g., pulsars, can be used to probe the strength of the parallel to the line of light magnetic field component in ionized media (see \cite{2005Sci...307.1610G}). The measurements are contaminated, however, by the poorly constrained variation of the thermal electron density. Similarly to synchrotron emission, dust particles get aligned with long axes perpendicular to the magnetic field due to the action of radiative torques (RATs) \citep{2007MNRAS.378..910L} and produce far-infrared emission polarization perpendicular to the magnetic field, while the polarization of starlight resulting from such an alignment is parallel to the magnetic field direction (\cite{Andersson2015} and ref. therein). This approach is applicable to dense media with sufficient dust density and it heavily relies on the alignment properties of dust that vary in space depending on the grain illumination, grain size distribution, and grain disruption \citep{2019arXiv191012205H,2021ApJ...908...12L,2022FrASS...9.3927T}. Additional techniques rely on the polarization that arises from Goldrech-Kylafis effect \citep{1981ApJ...243L..75G,2010MNRAS.402L..64C} and atomic alignment \citep{YL06,YL10, ZhangHCL2020}. Their applications have been limited so far. 

All techniques above employ polarization. However, it is known that high-precision polarization measurement is challenging and requires significantly more effort than measurements of signal intensities. As a result, the introduction of a new technique, namely, the Gradient Technique (GT), which may get the magnetic field information without polarization measurements, opens a new avenue for studying astrophysical magnetic fields in diffuse media.\footnote{Note that GT can also employ polarization to get extra information about the magnetic field. For instance, as shown in \cite{LY18b}, Synchrotron Polarization Gradients (SPGs) can use synchrotron polarization at different wavelengths to probe magnetic fields at different distances along the line of sight \citep{Ho19,ZhangLazarian2019,ZhangHCL2020}, while Faraday Gradients (FGs) can get the distribution of plane of sky direction of the magnetic field. However, we do not discuss polarization versions in the present paper.} The GT employs the properties of magnetic turbulence. In this paper, our primary goal is to explore the properties of {\it Synchrotron Intensity Gradients (SIGs)} \citep{LY17a} that employ gradients of synchrotron intensities. However, our results apply to other incarnations of the GT, e.g., to {\it Velocity Gradient Technique (VGT)} with its subdivision of Velocity Centroid Gradients (VCGs) that employ Velocity Centroids \citep{GL17,YL17a} and Velocity Channel Gradients (VChGs) \citep{LY18a} that employ intensity fluctuations in thin channel maps. For subsonic turbulence, density fluctuations arising as entropy fluctuations mimic velocity fluctuations \citep{Davidson2015Turbulence}. 

Therefore, the density gradients can be used as proxies of velocity gradients \citep{2019ApJ...886...17H}. Keeping in mind the application of the technique to diffuse media in galaxy clusters, i.e., Intra Cluster Media (ICM), we study the application of Intensity Gradients (IGs) to X-ray emissivity of turbulent medium.

The GT has been successfully applied to galactic and extragalactic environments with the magnetic maps obtained with the GT successfully compared to those obtained with polarization \citep{2019ApJ...886...17H,2021ApJ...911...37H,2022ApJ...941...92H,Hu2024Nature}.  It was also applied to intergalactic media with GT results compared with the magnetic field structure obtained with numerical simulations of galaxy cluster formation \citep{2020ApJ...901..162H}. The latter case presents is of special interest as the turbulence in galaxy clusters is super-{\toreferee Alfv{\'e}nic}, i.e. has the {\toreferee Alfv{\'e}n} Mach number $M_A=V_L/V_A>1$, {\toreferee where $V_L$ represents the injection velocity}. At the same time, most numerical testing of GT was performed for $M_A \leq 1$. 

The present study aims to explore the GT's ability to map magnetic fields in environments corresponding to $M_A> 1$ and $ M_A \gg 1$. This study's primary astrophysical application is related to justifying magnetic field studies with the GT in the ICM, but it is also applicable to other branches of the GT.


Below, this paper will utilize the numerical simulation to study the anisotropy statistics of the super-{\toreferee Alfv{\'e}nic} fluid on both small and large scales. Furthermore, we will explore the application of magnetic field tracing with the gradient technique. In what follows we structure our paper in the following way. Section \ref{sec:theory} will cover the theoretical considerations behind the GT approach to mapping magnetic fields. In section \ref{sec:method}, we will cover the detail of the numerical method.
In contrast, in section \ref{sec:result}, we will discuss the application of synchrotron intensity gradients (SIGs) to {\toreferee super-Alfv{\'e}nic} turbulence and provide a comparison with the sub-{\toreferee Alfv{\'e}nic} case. Section \ref{sec:result2} is devoted to effects noise and compressibility. It also illustrates the capabilities of GT in magnetic field tracing with X-ray data. In Section \ref{sec:discussion}, we will discuss the applicability of the results and further research directions. Section \ref{sec:summary} will summarize the results of the paper.

\section{Theoretical Consideration}
\label{sec:theory}
\begin{figure*}
\includegraphics[width=0.62\paperheight]{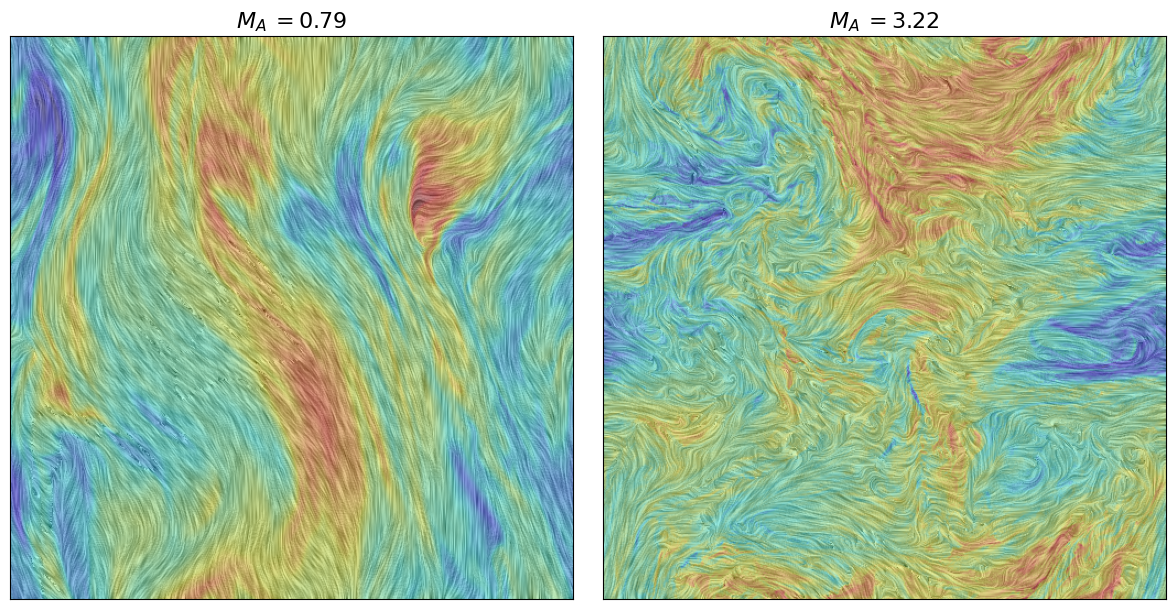}
\caption{Intensity of projected velocity fluctuation (color plot) overlaid with projected magnetic field lines (streamline plot). }
\label{fig:moro}
\end{figure*}
\subsection{sub-{\toreferee Alfv{\'e}nic} Turbulence and GT}

The level of turbulence controls the magnetic field structure and properties of magnetic turbulence. For sub-{\toreferee Alfv{\'e}nic} and trans-{\toreferee Alfv{\'e}nic} turbulence, the observational line-of-sight averaged gradients are dominated by the  dominated by {\toreferee Alfv{\'e}nic} modes.

The {\toreferee Alfv{\'e}nic} Mach Number $ M_A$ can quantify the magnetization level. Depending on different astrophysical objects, $M_A$ varies from the sub-{\toreferee Alfv{\'e}nic} regime, i.e., $M_A<1$ the super-{\toreferee Alfv{\'e}nic} regime, i.e., $M_A>1$ with the special regime of trans-{\toreferee Alfv{\'e}nic}, i.e., $M_A=1$ regime in between. The hot phase of the interstellar medium (ISM) (see \cite{2006ApJ...636.1114D} for a list of the idealized ISM phases) $M_A<1$. The turbulence may be both sub-{\toreferee Alfv{\'e}nic} and super-{\toreferee Alfv{\'e}nic} in molecular clouds and super-{\toreferee Alfv{\'e}nic} in an intracluster medium (ICM). 
Graphically, the level of magnetization is associated with the topology of the field lines is shown in Figure \ref{fig:moro}. 
The streamlines represent the B-field direction. The left panel of the figure depicts B-field lines with only minor variations in the sub-{\toreferee Alfv{\'e}nic} regime. In contrast, the right panel of Figure \ref{fig:moro} shows a highly disorganized magnetic field in the super-{\toreferee Alfv{\'e}nic} case.
Naturally, the difference in magnetic field structure can entail the difference in the properties of gradients. Thus, below, we briefly explain what is known about the magnetic field properties in sub-{\toreferee Alfv{\'e}nic} and super-{\toreferee Alfv{\'e}nic} regimes.

The magnetization level could also shape the local fluctuation on other physical quantities and can be used to trace the B-field lines. Modern Theory of MHD turbulence 
 (see \cite{BL19}) suggests that for $M_A\le 1$ a scale-dependent anisotropic cascade exists of {\toreferee Alfv{\'e}nic} modes. This cascades shapes the cascade of slow modes while leaving fast MHD modes marginally affected (\cite{GS95, LV99}, GS95 and LV99, respectively, \cite{CL02,CL03,2001ApJ...562..279L}). For {\toreferee Alfv{\'e}n} and slow modes, the eddies are elongated along mean field lines, and the elongation increases at a small scale. The GT, when applied to {\toreferee sub-Alfv{\'e}nic} turbulence, utilizes this anisotropy to trace the magnetic field. The physics of this tracing is obvious from the picture of magnetic eddies aligned with the magnetic field that follows from the theory of turbulent reconnection in LV99. Indeed, LV99 predicts that the time scale for turbulent reconnection equals the eddy turnover time. Therefore, the eddies that mix magnetized fluid in the direction perpendicular {\it local} magnetic field that surrounds the eddies are not subject to the magnetic field back-reaction. This provides the natural direction of least resistance along which the turbulent energy cascades. As the eddies mix the magnetized fluid, the {\toreferee Alfv{\'e}nic} perturbations are induced within a period equal to the eddy turnover time. Equating the two values, one gets 
 \begin{equation}
   l_\bot/v_l = l_\parallel/V_A, 
   \label{balance}
\end{equation}
where $l_\parallel$ is the eddy extent parallel to the local direction of the magnetic field, while $l_\bot$ is the eddy extent perpendicular to the local magnetic field.  It was shown in LV99 that for $M_A\le 1$ Eq. (\ref{balance}) entails the relation between the parallel and perpendicular eddy sizes
\begin{equation}
l_{\|}\approx L \left(\frac{l_\bot}{L}\right)^{2/3} M_A^{-4/3},
\label{lpar}
\end{equation}
where $L$ is the injection scale of turbulence. Eq. (\ref{lpar}) testifies that the eddies get more and more elongated along the local directions of the magnetic field as the eddy size decreases. Thus, at a sufficiently small scale, eddies act as compass needles aligned with the nearby magnetic fields. Naturally, the gradients of both velocities and magnetic fields arising from such eddies are perpendicular to the local magnetic field. In other words, the velocity and magnetic field gradients can act similarly to dust grains aligned by the magnetic field and trace the field direction. 

Eq. (\ref{balance}) is known as {\it critical balance}. This concept was introduced in the pioneering MHD turbulence study in GS95. However, as a note of caution, we should mention that in GS95, the critical balance is formulated in Fourier space with the parallel to magnetic field axis aligned with the mean magnetic field. Similarly, for $M_A=1$, GS95 formulated the relation for the scale-dependent anisotropy $k_\|\sim k_\bot^{2/3}$ in the mean magnetic field system of reference. The corresponding confusion in relation to the system of references still exists in the literature, although the fact that the {\toreferee Alfv{\'e}nic} turbulence scaling is only valid in the system of reference was confirmed by numerical simulations \citep{CV00,MG01} and is the basis of the further theoretical studies \citep{YL02, 2004ApJ...604..671F}. The notion of {\it local} system of reference is absolutely fundamental for the GT. 

For $M_A\le 1$, Eq. (\ref{balance}) is satisfied for scales smaller than the scale (see LV99)
\begin{equation}
    l_{trans}\approx L M_A^2.
\end{equation}
For scales in the range $l_{trans}<l<L$, the turbulence isotropically injected at scale $L$ is in the weak regime, in which the cascade induces the decrease of $l_\bot$ which keeps the parallel scale of turbulent fluctuations unchanged (LV99, \cite{2000JPlPh..63..447G}). As a result, in both regimes the gradients are perpendicular to the magnetic field direction and GT can trace magnetic field. 

For $M_A\leq 1$ the scaling of turbulent velocity motions at $l<l_\text{tran}$ (LV99)
\begin{equation}
v_l\approx V_L \left(\frac{l_\bot}{L}\right)^{1/3} M_A^{1/3},
\label{vAlf}
\end{equation}
i.e., Kolmogorov if the velocity is measured as a function of scale $l_\bot$, i.e., $v_l\sim l_\bot^{1/3}$. Naturally, the same type of scaling, i.e., $b_l \sim l_\bot^{1/3}$, is expected for {\toreferee Alfv{\'e}nic} turbulent fluctuations. As we mentioned earlier, the slow modes are slaved by {\toreferee Alfv{\'e}n} modes and copy the {\toreferee Alfv{\'e}n} mode scaling. As a result, both for {\toreferee Alfv{\'e}n} and slow modes, the gradients scale as 
\begin{equation}
    v_l/l_\bot \sim b_l/l_\bot \sim l_\bot^{-2/3}
    \label{grad}
\end{equation}
 i.e., the turbulent motions at the smallest scales induce the largest amplitude gradients. Together with the fact that the gradients are closely aligned with {\it local} magnetic field, this enables mapping of projected magnetic fields.


 \subsection{super-{\toreferee Alfv{\'e}nic} turbulence and GT}

The turbulence is different for super-{\toreferee Alfv{\'e}nic} and sub-{\toreferee Alfv{\'e}nic} cases. This can be seen from Figure \ref{fig:moro}, which shows the synthetic observations of magnetic field lines in turbulence corresponding to $M_A<1$ (left panel) and $M_A>1$ (right panel).

It is obvious that if $M_A\gg 1$, most of the turbulent energy is in the form of hydrodynamic motions. For our purposes, {\toreferee the action of the non-linear turbulent dynamo} can be disregarded as it transfers only $3/38$ fraction of the energy cascade into the magnetic field \citep{XL16}. Therefore, in the weakly compressible limit, i.e., for low sonic Mach number, $M_s=V_L/V_s$, where $V_s$ is the sound velocity, we expect to have the Kolmogorov cascade for supersonic turbulence at the injection scale. 

In Kolmogorov turbulence, turbulent motions scale as $v_l\sim (l/L)^{1/3}$. Thus, the kinetic energy of turbulent eddies decreases as $(l/L)^{2/3}$ with the decrease of the scale $l$. As a result for turbulence with $M_A>1$ at the transition scale \citep{L06}
\begin{equation}
    l_{A}=L M_A^{-3}
    \label{eq:lA}
\end{equation}
the turbulent velocity gets equal to the {\toreferee Alfv{\'e}n} velocity $V_A$. 

Therefore, one can roughly subdivide the range of turbulent scales into two regions: Kolmogorov turbulence for $l_{tr}<l<L$ and the MHD trans-{\toreferee Alfv{\'e}nic turbulence for $l<l_{tr}$. For the first regime where at these large scales ($l > l_A$), the turbulence is isotropic, magnetic fields are passively moved by strong hydrodynamic motions. For the second regime, magnetic fields are dynamically important and determine the evolution of the cascade \citep{GS95,Kraichnan1965}. One can expect that the GT will act differently for these two regimes.}   

\subsection{Gradients and Plane of Sky Magnetic Field Direction}

The measurements of 3D magnetic fields with gradients are rarely available. As a rule, one gets the measurements that include both line-of-sight averaging and telescope beam averaging. Due to the latter effect, the smallest scale at which the gradients can be measured is determined by the telescope resolution $l_{res}$. Therefore, according to Eq. (\ref{grad}), the GT samples magnetic field structure at scales larger than $l_{res}$. 

The averaging along the line of sight further modifies the gradient statistics. The gradient measured from the 2D observation maps can be considered the proxy of the 3D fluctuation \citep{LY17a}. Gradients are measured for turbulent volume extended by $\mathcal{L}>L_{inj}$ along the LOS, and this entails additional complications, where $\mathcal{L}, L_{inj}$is the LOS depth and the injection scale. While eddies stay aligned with the local magnetic field, the direction of the local magnetic field is expected to change along the LOS. Thus, the contribution of the 3D velocity gradient is also summed up along the line of sight.

The summation of gradients, however, is different for sub-{\toreferee Alfv{\'e}nic} and super-{\toreferee Alfv{\'e}nic} cases. The contributions of gradients at the scale $l_{res}$ are being summed up along the line of sight. For $M_A<1$ and line of sight perpendicular to the mean magnetic field, the gradients add up along the line of sight, reflecting the structure of the averaged magnetic field along the line of sight. The same structure is being sampled by velocity and magnetic field gradients due to Eq. (\ref{grad}).

For $M_A>1$, consider first velocity gradients. If the scale $l_{res}< l_A$, the contributions from regions of the size $l_A$ are summed up along the line of sight. These contributions aligned with the mean magnetic field in independent magnetic domains of size $l_A$. The latter is randomly oriented; thus, the summation is happening in a random walk manner. This increases the role of the gradients arising from the large-scale motions of the largest eddies $\sim L$. Finally, for $l_{res}\gg l_A$, the velocity gradients sample only hydrodynamic eddies. The contributions from the gradients arising from eddies at the scale  $l_{res}$ are completely random and cancel out due to line-of-sight averaging. 

The magnetic field for $M_A\gg 1$ is passively advected by velocity motions on the scales much larger than $l_{res}$. Therefore, the magnetic fields are aligned with the flow lines of the large-scale eddies. The velocity gradients are perpendicular to the flow lines in solenoidal hydrodynamic turbulence. This results in the observed velocity gradients being perpendicular to the magnetic fields in large-scale eddies. As for $M_A\gg 1$, magnetic field lines follow the velocity flow lines; the magnetic field gradients are also perpendicular to the direction of the large-scale magnetic field.  

In the present study, we will analyze velocities obtained with velocity centroids and magnetic fields represented by synchrotron intensities. A limited study of synchrotron gradients for $M_A>1$ is presented in \cite{LY17a}, but here we provide an extensive parameter study of this regime. {\toreferee We also note that this study only considers the case of magnetic fields in driven turbulence, and the results may differ in more complex astrophysical environments.}

\section{Numerical Method}
 \label{sec:method}

Motivated by studies of magnetic fields in the ICM medium, we focus on weakly compressible media corresponding to small sonic $M_s=V_L/V_s$, where $V_s$ is the sound velocity and high {\toreferee Alfv{\'e}n} Mach $M_A$ numbers. Therefore, we employ super-{\toreferee Alfv{\'e}nic} incompressible simulations formally corresponding to $M_s=0$. To explore the effects of compressibility, however, we compare our results with super-{\toreferee Alfv{\'e}nic} compressible simulations with $M_s\sim 1$.

The set of simulations is chosen so that for the low {\toreferee Alfv{\'e}n} Mach number $M_A\sim 2$, the {\toreferee Alfv{\'e}nic} scale $l_A$ (see Eq. \ref{eq:lA}) is resolved, while for the largest {\toreferee Alfv{\'e}n} Mach number $M_A\sim 5$, the entire simulation corresponds to super-{\toreferee Alfv{\'e}nic} regime, i.e. $l_A$ is less than the scale of numerical dissipation (see table \ref{table:simulation}, and  table \ref{table:Athena++} )

\subsection{Super-{\toreferee Alfv{\'e}nic} Incompressible Simulations}
The gird-based method (e.g., Finite Volume/Difference) often introduces artificial viscosity, marking the fluid to enter the dissipation regime soon and hard to study the anisotropy statistics of MHD turbulence in a small scale. So, we use the pseudo-spectral method to simulate super-{\toreferee Alfv{\'e}n} fluid in this paper. The simulations were performed using the pseudo-spectral code \texttt{MHDFlows.jl} \citep{MHDFlows}\footnote{https://github.com/MHDFlows/MHDFlows.jl}. \texttt{MHDFlows.jl} is the newly developed MHD code based on the dynamical language Julia with \texttt{FourierFlows.jl}\citep{FourierFlows} framework. In contrast to the traditional spectral solver, it supports native GPU acceleration. In our paper, we solve the ideal incompressible MHD equation in the periodic box with the size of $2\pi$:

\begin{equation}
\begin{aligned}
\frac{\partial \vec{v} }{\partial t} + (\vec{v} \cdot \nabla )\vec{v} = -\nabla P + (\nabla \times \vec{B}) \times \vec{B} + \nu \nabla^2 \vec{v} 
 \\
\frac{\partial \vec{B} }{\partial t} = \nabla\times(\vec{v}\times\vec{B}) + \mu \nabla^2 \vec{B}
\end{aligned}
\end{equation}
All the symbols have their usual meaning. Pressure P is chosen such that the equations maintain the divergence-free condition throughout the simulation. For super-{\toreferee Alfv{\'e}nic} fluid in astrophysics, it is often under the sub-sonic regime, meaning that the compressibility of fluid is a weak and incompressible simulation is an adequate choice to study the behavior of the fluid. For each simulation, The turbulence is driven on large scale through the method proposed by \cite{A99}. In addition, a weak seed field was injected at the beginning of the simulation, and we analyzed the result after three large-scale eddies turnover times. We choose 2 Storage 5 Stages RK4 method (LSRK54) (see \cite{LSRK54} for theory and  \cite{Fletcher2015} for actual Implementation) for the time integration and 2/3 alias rule. Table \ref{table:simulation} shows the key parameter of the simulation.  
\begin{table}
\begin{tabular}{@{}lllll@{}}
\toprule
Simulation & $M_A$ & $N^3$                    & $\mu\:or\:\eta$                      & $l_A/\Delta x$ \\ \midrule
M1         & 2.4   & \multirow{5}{*}{$512^3$} & \multirow{5}{*}{$5\times10^{-5}$} & 37.0           \\ \cmidrule(r){1-2} \cmidrule(l){5-5} 
M2         & 2.9   &                          &                                   & 21.0           \\ \cmidrule(r){1-2} \cmidrule(l){5-5} 
M3         & 3.2   &                          &                                   & 15.6           \\ \cmidrule(r){1-2} \cmidrule(l){5-5} 
M4         & 5.2   &                          &                                   & 3.64           \\ \cmidrule(r){1-2} \cmidrule(l){5-5} 
M5         & 7.8   &                          &                                   & 1.07           \\ \bottomrule
\end{tabular}
\caption{Key parameters of the MHD incompressible simulation that are used in this paper.$M_A,\mu/\eta,N^3$ refer to the {\toreferee Alfv{\'e}nic} Mach Number, kinetic/magnetic diffusivity, and resolution of the simulation. The term "$l_A/\Delta x$" refers to the ratio between the transition scale and the pixel resolution. When $l_A/\Delta x>>1$, it indicates that the simulation is in a well-resolved regime, while if $l_A/\Delta x$ is close to 1, it means that the simulation is in a barely resolved regime.}
\label{table:simulation}
\end{table}

\subsection{ super-{\toreferee Alfv{\'e}nic} Compressible Simulation}
To gain insight into the impact of compressibility on the results, we include two super-{\toreferee Alfv{\'e}nic} trans-sonic simulations in our paper. We employ the 3D MHD simulations generated from the Athena++ MHD code \citep{Athena++} to set up a 3D, uniform, and isothermal turbulent medium. The simulations are set up with periodic boundary conditions with solenoidal turbulence injections.

In order to retain the small-scale spatial information required for the GT, we select the $4^{th}$ order reconstruction method from Athena++. After two large-scale eddy turnover times, a snapshot of the simulation is analyzed. Table \ref{table:Athena++} shows the key parameter of the simulation.  

\begin{table}
\begin{tabular}{@{}lllll@{}}
\toprule
Simulation & $M_S$ & $M_A$ & Resolution               & $l_A/\Delta x$ \\ \midrule
A1        & 1.04  & 2.3   & \multirow{2}{*}{$512^3$} & 42.1           \\ \cmidrule(r){1-3}
A2        & 1.02  & 4.8   &                          & 4.62           \\ \bottomrule
\end{tabular}
\caption{Key parameters of the Athena++ simulation that are used in this paper.}
\label{table:Athena++} 
\end{table}

\section{Synchrotron Intensity Gradients}
\label{sec:result}

\subsection{SIGs for super-{\toreferee Alfv{\'e}nic} turbulence}

For the power-law distribution of electrons \( N(E) E \sim E^{\alpha} dE \), the synchrotron emissivity is
\begin{equation}
\begin{aligned}
I_{sync}({\bf X}) \propto \int dz B^{\gamma}_{POS} (\bf X,z)
\end{aligned}
 \end{equation}
where $B^{\gamma}_{POS} = \sqrt{B_x^2+B_y^2}$ corresponds to the magnetic field component perpendicular to the line of sight, ${\bf X}$ is the plane of sky vector defined in x and y direction, z the line of sight axis and, $B_x,B_y$ the 3D magnetic field in x and y direction. {\toreferee We consider that the variations of cosmic ray density $n_{cr}$ occur on scales larger than those of the magnetic field variations, so that the synchrotron intensity depends only on the magnetic field.}
{\toreferee The fractional power of the index $\gamma = (\alpha+1)/2$ was an impediment for quantitative synchrotron statistical studies. As discussed in \cite{LP12}, the choice of $\gamma$ varies depending on which physical processes play a role in shaping the spectra, including shock acceleration, turbulence reacceleration, and propagation. These processes lead to a range of $\gamma$ values, typically between 1 and 4. \cite{LP12} showed that the correlation functions and spectra of $B^{\gamma}_{\bot}$ could be expressed as $\alpha = 3$, which gives $\gamma$ and therefore the dependence of synchrotron intensity on the squared magnetic field strength. We pick $\gamma = 2$ as it is similar to the case of observed cosmic-ray index $\alpha \approx 2.7$. In \cite{LP12}, the relation between the structure functions obtained for different $\gamma$ of synchrotron emission and those obtained for $\gamma=2$ was established.}

\subsubsection{Gradients}
This section will briefly explain the procedure of applying gradients and our {\toreferee analysis}. 


Due to the statistical nature of turbulence, the individual gradient vector orientation may not represent the local magnetic field direction. Therefore, one should use the statistical distribution of gradients to trace the magnetic field. The finding converts to the technique called sub-block averaging \cite{YL17a}. 

We divide the observational map into different sub-regions to trace the gradient orientation in a sub-region. For each sub-region, we conduct orientation statistics of gradient vectors and find their best fit of the Gaussian profile, in which the peak of the Gaussian profile reflects the statistical most probable magnetic field orientation in this sub–block. One important note is that as the area of the sampled region increases, the magnetic field's prediction traced through Gaussian block fit becomes more and more accurate. This means that to increase the accuracy of gradients, we have to sacrifice some of the resolution of the resulting magnetic field maps. 

To quantify how good gradients and magnetic fields are aligned, we employ the {\it alignment measure} AM that is introduced in analogy with the grain alignment studies (see \citealt{L07}):
\begin{equation}
\label{eq:AM}
AM=2\langle\cos^2\theta_r\rangle-1,
\end{equation}
and was discussed for the GT in \citealt{GL17,YL17a}).  The range of AM is $[-1,1]$ measuring the relative alignment between the {\it $90^o$-rotated} gradients and magnetic fields, where $\theta_r$ is the relative angle between the two vectors. A perfect alignment gives $AM=1$,  whereas random orientations generate $AM=0$, and a perfect perpendicular alignment, i.e., "wrong alignment case," corresponds to $AM=-1$. In what follows, we use $AM$ to quantify the alignments of GT with respect to the magnetic field.

\subsection{Magnetic field tracing in super-{\toreferee Alfv{\'e}nic}  observations}
\label{sec:result}
\begin{figure}
\includegraphics[width=0.26\paperheight]{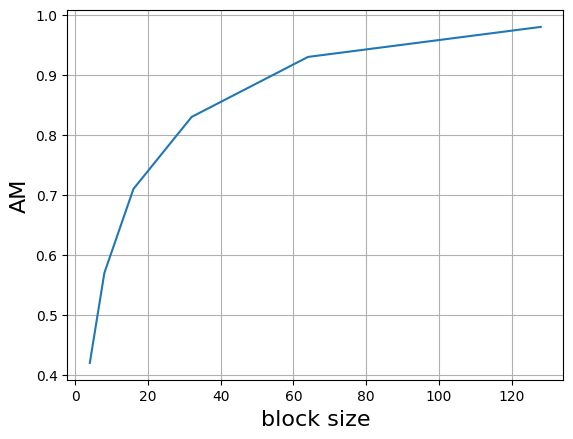}
\caption{The AM of gradient versus the block size using synchrotron intensity. 
The simulation used: M1 \\
Block size covered: [4,8,16,32,64,128] }
\label{fig:AM_block}
\end{figure}

\begin{figure}
\includegraphics[width=0.26\paperheight]{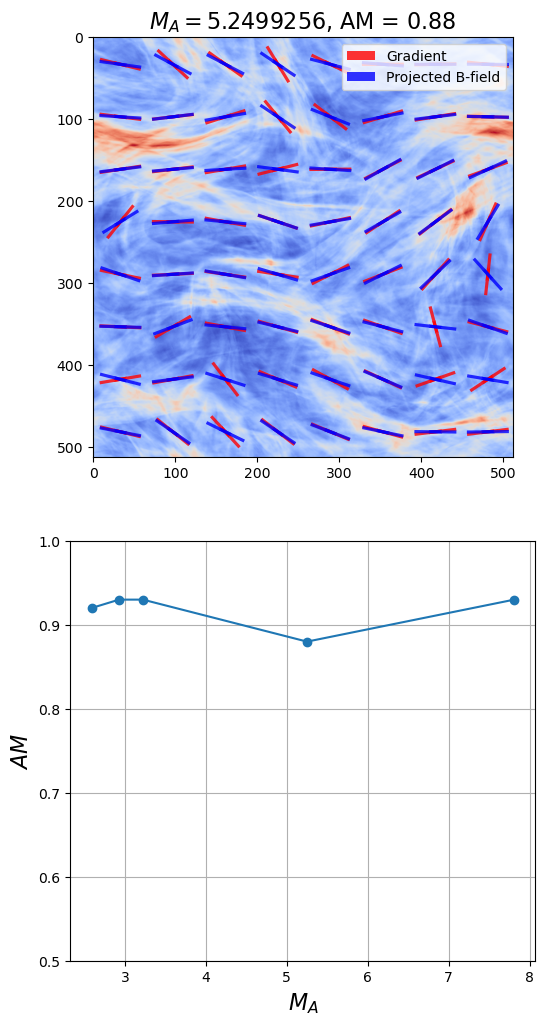}
\caption{Top panel: synchrotron intensity Map for simulation M4 overlay with gradient and polarization vector. Color with warmer colors represents stronger intensity.
Bottom panel: The AM across different $M_A$.}
\label{fig:AM}
\end{figure}
We construct a synthetic observation for synchrotron intensity (the method described at sec. \ref{sec:method} to study the gradient performance for simulations with {\toreferee Alfv{\'e}nic} Mach number $M_A$. We apply the sub-block averaging to both gradient and magnetic field vectors to trace the local gradient and magnetic field direction. Similar to studies in \cite{HL23a}, the accuracy of the magnetic field tracing is highly related to the choice of block size. The same applies to our super-{\toreferee Alfv{\'e}nic} simulations, and Figure \ref{fig:AM_block} shows the correlation between the two. We use AM to represent the statistical alignment between the gradient and magnetic field directions. The AM increase significantly from 0.4 to 0.93 when the block size is increased from $4^2$ to $64^2$. This supports our point that for small averaging, the gradients are unreliable, especially as the size of 4 points is below the numerical dissipation scale of our simulations. For better statistics, we fixed our paper's block size to $64^2$. Note that our studies are affected by numerical dissipation, and the smaller block sizes can potentially be applied for realistic observations for which the dissipation scale is much smaller than the resolution scale $l_{res}$.

Below we study the gradient technique in different {\toreferee Alfv{\'e}nic} Mach numbers with the fixed block size $64^2$. The top panel of Figure. \ref{fig:AM} shows a graphical result of applying the block averaging method to one of the simulations. The bottom panel shows the AM value for each simulation using the same block-averaging method. One can see from the top panel that, for the most part of the map, the calculated gradient vectors align well with the projected 2D B-field directions (red and blue arrow). It is only a few gradient vectors that are misaligned. As a result, we conclude that statistically the gradient vectors align well with B-field for both sub-{\toreferee Alfv{\'e}nic} and super-{\toreferee Alfv{\'e}nic} simulations. These trends apply to all of our super-{\toreferee Alfv{\'e}nic} simulations with an average $AM \approx 0.9$ (bottom panel, equivalent  to $\Delta \theta \sim \pm 10  ^{\circ}$).  We also note that the alignment only weakly depends on $M_A$. We note that such similar tendencies are also observed for compressible sub-{\toreferee Alfv{\'e}nic} simulation in \cite{HL23a} 

We also noted from Figure \ref{fig:AM_block} that part of the gradient vectors on the right-hand side diverges from the B-field direction. For regions with poor alignment, we observe that,

\begin{enumerate}

  \item  the amplitude of the plane of the sky magnetic field in those regions has a uniform angular distribution along the line of sight, meaning that the projected magnetic field has low amplitude 

  \item the gradient amplitude value of those regions tends to be low.

\end{enumerate}


\begin{figure}
\includegraphics[width=0.26\paperheight]{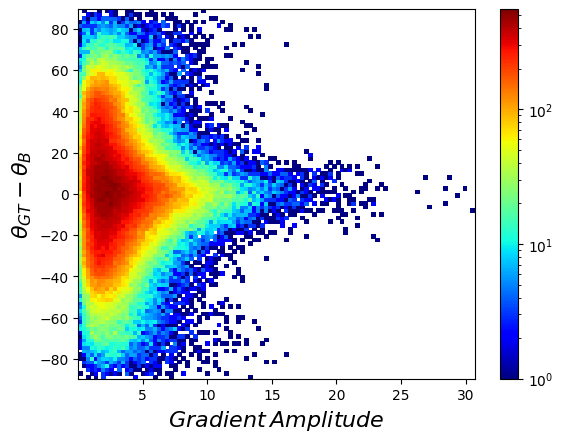}
\caption{The relative degree - gradient amplitude histogram. Y-axis: the relative degree between individual gradient vector directions. X-axis: the gradient amplitude value. Color bar: log scale pixel count. The simulation used: M4 }
\label{fig: phi_gradient}
\end{figure}

This means that the statistical significance of the magnetic field detection along such directions is low with polarization and gradients. Therefore, the significance of the observed discrepancies between the actual projected magnetic field and that measured with gradients is low. Such points will be within the noise.   

To test our hypothesis that in the super-{\toreferee Alfv{\'e}nic} case, the dispersion in angle between the gradient and the projected magnetic field is correlated with the gradient amplitude, we plot in Figure \ref{fig: phi_gradient}, a 2D histogram of relative angle versus gradient amplitude. We see a clear statistical relationship: the pixels with higher gradient amplitude are more aligned with the projected B-field, while the lower amplitude gradients have more dispersion. This finding opens a way of improving the tracing of the magnetic field by filtering out the points corresponding to low amplitudes of gradients. To improve the accuracy of GT magnetic field mapping, we will explore this and other types of filtering elsewhere.  


\subsection{Comparison: SIGs for sub-{\toreferee Alfv{\'e}nic} turbulence}

\begin{figure}
\includegraphics[width=0.32\paperheight]{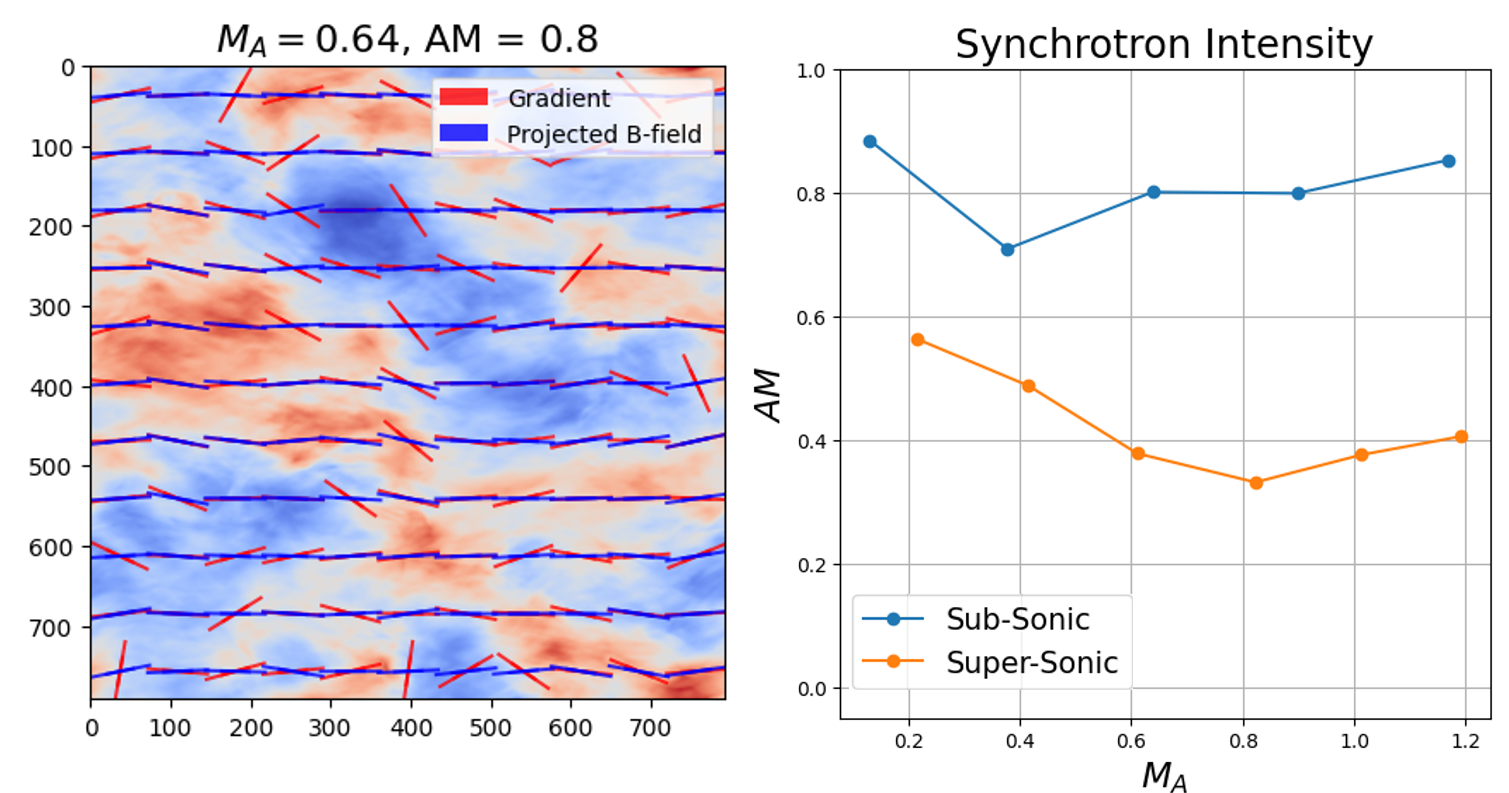}
\caption{Extracted Figure from \citep{HL23a}. The left panel shows the result of SIG in a subsonic sub-{\toreferee Alfv{\'e}nic} simulation. The right panel shows the change of AM in difference $M_A$ in both subsonic ($M_S \approx 0.6$) and supersonic($M_S\approx 6$)  regime }
\label{fig: New_Ms}
\end{figure}

The regime of sub-{\toreferee Alfv{\'e}nic} turbulence is very different regarding the physics underlying the gradient technique. To have a complete picture of the SIG performance, we present the results for sub-{\toreferee Alfv{\'e}nic} simulations. 

SIGs were introduced in \cite{LY17a} and applied to Planck synchrotron emission studies there. 
The left panel in Figure \ref{fig: New_Ms} shows the comparison of predicted directions of obtained with SIGs applied to data of subAlvenic sub-sonic simulations with the maps of the line of sight (LOS) projected magnetic field. The relationship between AM and $M_A$ in sub-{\toreferee Alfv{\'e}nic} simulations is displayed in the right panel of Figure \ref{fig: New_Ms}. This graph is extracted from the \cite{HL23a} and includes results from both sub-sonic and super-sonic regimes.

Comparing results in Figure \ref{fig: New_Ms} to Figure \ref{fig:AM} and Figure \ref{fig:athena++}, we conclude that the alignment is not sensitive to the magnetization and maintaining a good tracing performance.

\section{Additional effects}
\label{sec:result2}

\subsection{Synthetic observations: the presence of noise}
\begin{figure}
\includegraphics[width=0.26\paperheight]{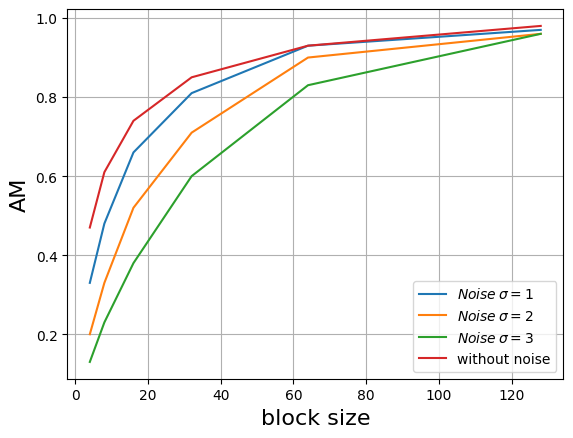}
\caption{Effect of noise for the gradient technique. Simulation used: M1}
\label{fig:noise}
\end{figure}
To test the impact of noise on the tracing performance of the gradient technique, we constructed a synthetic maps incorporating the effect of noise. The results are shown in Figure \ref{fig:noise}, which demonstrates the dependence of  the AM on the noise level.

Three different noise levels were added to the data, ranging from $1\sigma$ to $3\sigma$ of the map intensity value. 
The results indicate that AM decreases as the noise level increases. The noise impacts the performance more for small block sizes, which do not have sufficient statistics to start with, while the decline is milder for block sizes greater than $64^2$.

\subsection{Effect of compressibility}

Incompressible simulations provide a good representation of subsonic turbulence in various astrophysical media, e.g., in the ICM. To gauge the effect of compressibility for the GT, we employ our trans sonic simulations. The setup is described in \ref{sec:result}. 

\begin{figure}
\includegraphics[width=0.32\paperheight]{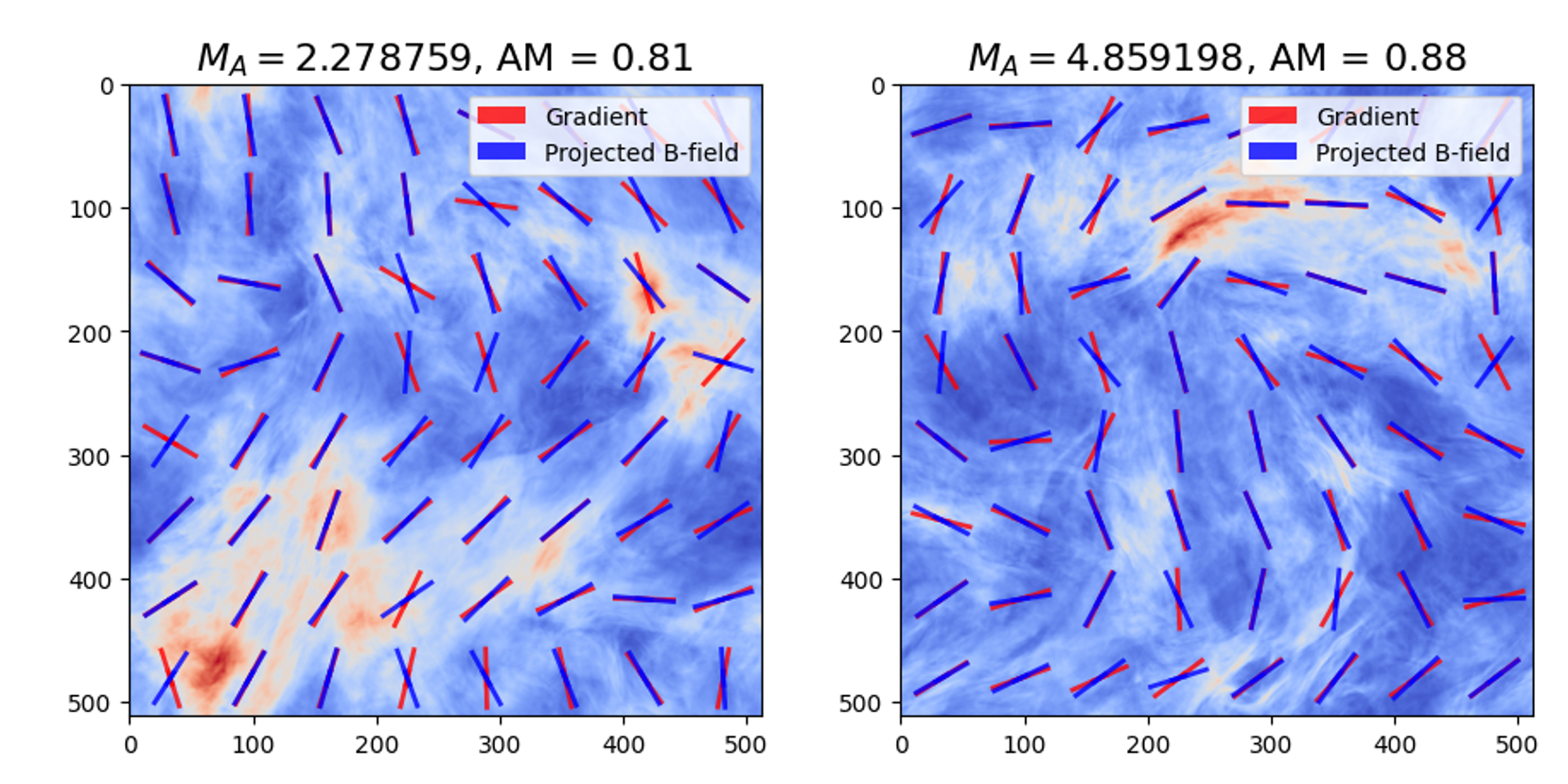}
\caption{Synchrotron intensity Map overlay with gradient and polarization vector. Simulation used : A1 (Left), A2(Right)}
\label{fig:athena++}
\end{figure}

Figure \ref{fig:athena++} demonstrates the magnetic field tracing by SIGs for trans-sonic super-{\toreferee Alfv{\'e}nic} turbulence. We see a decrease of $AM$, but one can see that even for this case, the general structure of the magnetic field can be correctly represented by the GT.  

Comparing GT results for trans-sonic super-{\toreferee Alfv{\'e}nic} turbulence in Figure \ref{fig:athena++} with those in Figure \ref{fig:AM}) that represent the subsonic super-{\toreferee Alfv{\'e}nic} case, we observe that the decrease of the AM is more noticeable for $M_A \sim 2$ compared to $M_A\sim 5$. The significance of this effect requires further studies, however.

In our sub-{\toreferee Alfv{\'e}nic} studies, we have developed a toolbox to improve the accuracy of GT \cite{HL23a,2022arXiv220806074H}. We expect a similar study for super-{\toreferee Alfv{\'e}nic} GT tracing to improve magnetic field mapping further.


\subsection{GT application to X-Ray Intensities}

\begin{figure}
\includegraphics[width=0.32\paperheight]{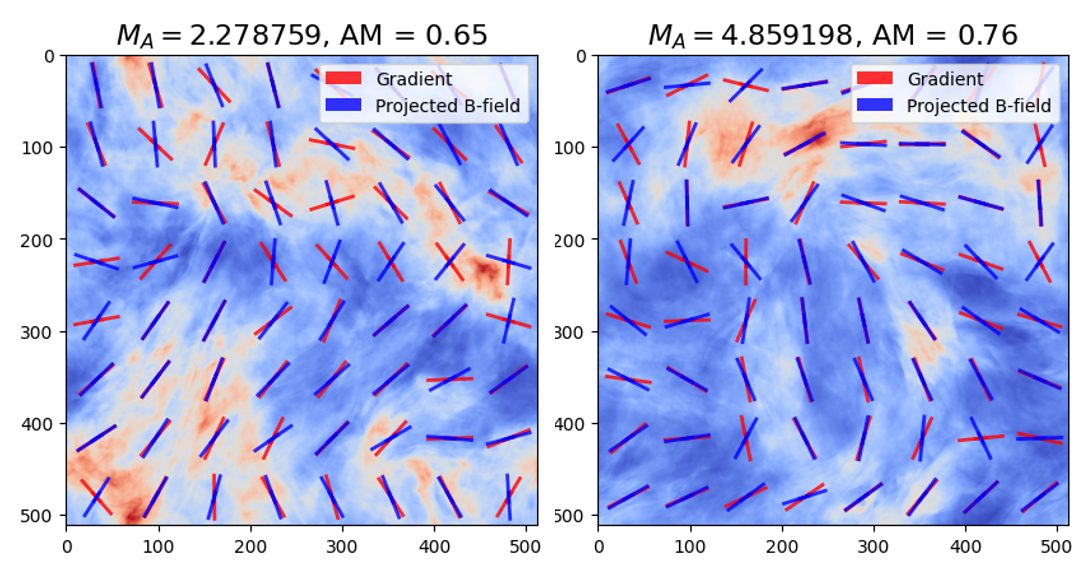}
\caption{X-ray intensity Map overlay with 
X-ray gradient and polarization vector. Simulation used : A1 (Left), A2(Right)}
\label{fig:XRay}
\end{figure}

Using X-ray maps, gradients were applied to map the magnetic fields in galaxy clusters \citep{2020ApJ...901..162H}. The point of whether the accuracy of such mapping depends on resolving the {\toreferee Alfv{\'e}nic} scale $l_A$ given by Eq. (\ref{eq:lA}) was not resolved in \cite{2020ApJ...901..162H}. Below we illustrate the effects of such mapping in settings that both marginally resolved $l_A$, i.e., corresponding to $M_A\sim 2$, and not resolved $l_A$, i.e., corresponding to $M_A\sim 5$. 

We construct a synthetic X-ray emission map using our Athena++ simulation. The intensity map, $I$, is defined as follows:

\begin{equation}
\label{eq:XRay}
I = \int_{LOS} n^2 dx.
\end{equation}

The result of the intensity gradient compared to the projected magnetic field is shown in Figure \ref{fig:XRay}. We observe that GT can be successfully applied to studying trans-sonic super-{\toreferee Alfv{\'e}nic} turbulence. Compared to Figure \ref{fig:athena++}, we observe a worse X-rays performance than synchrotron, which is expected, as for trans-sonic turbulence, the density statistics do not strictly follow that of velocities. However, the AM observed is significant enough to justify studying magnetic fields in turbulence for the case when $l_A$ is only marginally resolved or not resolved at all.


\section{Discussion}
\label{sec:discussion}
 Magnetic field tracing in galaxy clusters is a challenge. Observational studies using Faraday rotation showed the magnetic field line of clusters is weak (e.g., \cite{VE03, Bonafede2010}). Measuring the POS magnetic field component is very difficult \citep{Beck1996,Eatough2013,Pakmor2020} due to the Faraday depolarization of the signal.  In this situation, GT presents a unique way to map the magnetic fields in ICM.  
 
 While our earlier GT numerical studies mostly dealt with the turbulent medium in the sub-{\toreferee Alfv{\'e}nic} or trans-{\toreferee Alfv{\'e}nic} regime, the clusters of galaxies are in the super-{\toreferee Alfv{\'e}nic} regime. This deficiency is compensated for in this paper. In \cite{LY17a,Hu2024Nature}, the ability of SIGs to trace magnetic fields in super-{\toreferee Alfv{\'e}nic} turbulence was demonstrated for $M_A\approx 3$. In the current paper, we provide an extensive parameter study for different $M_A$ and prove that the magnetic field good magnetic field tracing is possible for various $M_A$. We prove that SIGs trace the magnetic field well both when the observations resolve and do not resolve the {\toreferee Alfv{\'e}n} scale $l_A$. This means we do not have rigid constraints on the required resolution of observations. 

The unique property of SIGs is that they are not sensitive to the Faraday depolarization effect, the latter being a serious limitation for present and future synchrotron polarization studies of ICM magnetic fields. Indeed, this depolarization is very strong for low-frequency synchrotron emission. However, the synchrotron emission from most of the ICM, e.g., from cluster halos, is observed at low frequencies only. Thus, SIGs provide a unique way to map magnetic fields in galaxy clusters and recently discovered larger structures, i.e., Megahaloes \cite{2022A&A...657A..56K}. This paper also provides additional support for the magnetic field maps obtained by \cite{Hu2024Nature}.
 
Our work shows that SIGs present a powerful tool for mapping super-{\toreferee Alfv{\'e}nic} magnetic fields in galaxy clusters but are not limited by this.
Our previous study in \cite{Hu2020} argued that the magnetic lines could be traced using the GT applied to X-rays. Applying the GT to X-ray maps was intriguing as it provides a way to utilize these maps for a new purpose, i.e., for magnetic field studies. This opened an avenue for significantly increasing the scientific output of the X-ray galactic cluster observations. However, in the earlier study, the resolution of the scale $l_A$ was assumed to be essential for this tracing. In reality, $l_A$ is not well defined, and thus, the accuracy of magnetic field mapping in \cite{Hu2020} could be questioned. The removal of the constraint of resolving $l_A$ that we demonstrated provides valuable support with the accuracy of maps obtained with X-ray maps in \cite{Hu2020}.



Our work is also important in a broader context of the advancement of GT technique. It has numerically demonstrated that the GT can accurately trace magnetic fields in various turbulence regimes, from $M_A<1$ to $M_A\gg 1$. While our paper only deals on the gradient in the synchrotron and X-ray intensity, the concept of gradient also applies to other data sets, e.g., velocity centroids and velocity channel data. In addition, magnetic field tomography using Synchrotron Polarization Gradients (SPGs) \cite{LY18b} was tested for sub-{\toreferee Alfv{\'e}nic} turbulence. Our present study suggests that the possibility of such a tomographic study can also apply to the super-{\toreferee Alfv{\'e}nic} case. {\toreferee Those results suggested the broad application of G.T. in different astrophysical environment, that may handle handle more complicated conditions, including contact discontinuities where fluid properties and magnetic field jump almost arbitrarily. These applications would be worth to be explored in the future study}

\section{Summary}
\label{sec:summary}
In this paper, we numerically studied the applicability of the GT for tracing magnetic field in super-{\toreferee Alfv{\'e}nic} turbulence. Our main results are:

1. Even though the turbulence is large-scale super-{\toreferee Alfv{\'e}nic} and the cascade is hydrodynamic, the GT can successfully trace the magnetic field. This results from dynamically unimportant magnetic fields passively moving by powerful large-scale hydrodynamic eddies.

2. The effects of noise and compressibility do not prevent POS magnetic tracing by the GT. The sub-block averaging procedure works reliably in the super-{\toreferee Alfv{\'e}nic} case.

3. The GT was demonstrated to apply to magnetic field tracing using synchrotron intensities and X-ray maps for conditions similar to those in clusters of galaxies.

\section*{Acknowledgments.} 
We acknowledge Ka Ho Yuen and Yue Hu for the fruitful discussions. We acknowledge the support the NASA ATP AAH7546 and NASA TCAN 144AAG1967 grants.

\vspace{5mm}

\software{Athena++ \citep{Athena++}, MatPlotLib \citep{2007Matplotlib}, Julia \citep{Julia}, MHDFlows \citep{MHDFlows}}

\bibliographystyle{aasjournal.bst}
\bibliography{ref}

\appendix
\begin{table}[h]
\centering
\caption{List of Acronyms}
\begin{tabular}{|l|l|}
\hline
\textbf{Acronym} & \textbf{Definition} \\
\hline
AM & Aligment Measrure \\
SIG & Synchrotron Intensity Gradients  \\
GT  & Gradient Techinque \\
ICM & Intracluster Medium \\
ISM & Interstellar Medium \\
LOS & Line of Sight (z)\\
MHD & Magnetohydrodynamic \\
POS & Plane of Sky (x-y)\\
GS95 & \cite{GS95} \\
LV99 & \cite{LV99} \\
\hline
\end{tabular}
\end{table}


\end{document}